\documentstyle[aps,prl,twocolumn]{revtex}

\newcommand{\beq}{\begin{equation}}
\newcommand{\eeq}{\end{equation}}
\def\beqa{\begin{eqnarray}}
\def\eeqa{\end{eqnarray}}

\def\lap{\lower.5ex\hbox{$\; \buildrel < \over \sim \;$}}
\def\gap{\lower.5ex\hbox{$\; \buildrel > \over \sim \;$}}

\begin{document}
\wideabs{
\title{Superconducting cosmic strings as Gamma Ray Burst engines}

\author{V. Berezinsky$^{1)}$, B. Hnatyk$^{1),2)}$ and A. Vilenkin$^{3)}$}

\address{$^{1)}$INFN, Laboratori Nazionali del Gran Sasso, I-67010 Assergi
(AQ), Italy \\
$^{2)}$Institute for Applied Problems in Mechanics and Mathematics,
NASU, Naukova 3b, Lviv-53, 290053, Ukraine\\
$^{3)}$Physics Department, Tufts University, Medford, MA 02155, USA}

\maketitle
\begin{abstract}
Cusps of superconducting strings can serve as GRB engines. A powerful
beamed pulse of electromagnetic radiation from a cusp produces a jet of
accelerated particles,
whose propagation is terminated by the shock responsible for
GRB. A single free parameter,
the string scale of symmetry breaking $\eta \sim  10^{14}~GeV$,
together with reasonable assumptions about the magnitude of
cosmic magnetic fields and the fraction of volume that
they occupy,
explains the GRB rate, duration
and fluence, as well as the observed ranges of these
quantities. The wiggles on the string can drive the short-time structures
of GRB.
This model predicts that GRBs are accompanied by strong bursts of
gravitational radiation which should be detectable by LIGO, VIRGO
and LISA detectors.
\end{abstract}

}

Models of gamma ray bursts (GRBs) face the problem of explaining the
tremendous energy released by the central engine \cite{Pir98}.
In the case of isotropic emission, the total energy output should be as
high as $4\times 10^{54}$ ergs. Strongly beamed emission
is needed for all known
engine models, such as mergers and hypernovae, but such extreme
beaming is difficult to arrange.
In this
paper we show that emission of pulsed electromagnetic radiation
from cusps of superconducting cosmic
strings naturally solves this problem and explains
the observational GRB data using only one engine parameter.

Cosmic strings are linear defects that could be formed at a symmetry
breaking phase transition in the early universe \cite{Book}.  Strings
predicted in most grand unified models respond to external
electromagnetic fields as thin superconducting wires \cite{Witten}.
As they move through cosmic magnetic fields, such strings develop
electric currents.
Oscillating loops of
superconducting string emit short bursts of highly beamed
electromagnetic radiation
and high-energy particles
\cite{VV,SPG}.

The idea that GRBs could be produced at cusps of
superconducting strings was first
suggested by Babul, Paczynski and Spergel \cite{BPS} (BPS) and further
explored by Paczynski \cite{Paczynski}.
They assumed that the bursts
originate at very high redshifts ($z\sim 100 - 1000$),  with GRB
photons produced either directly or in electromagnetic cascades
developing due to interaction with the microwave background.
This model requires the existence of a strong primordial magnetic
field to generate the string currents.

As it stands, the BPS model does not agree
with observations.  The observed GRB redshifts are in the range
$z\lesssim 3$, and the observed duration of the bursts ($10^{-2}s
\lesssim \tau\lesssim 10^3 s$) is significantly longer than that
predicted by the model.  On the theoretical side, our understanding of
cosmic string evolution  and of the GRB generation in relativistic jets
have considerably evolved since the BPS papers
were written.  Our goal in this paper is to revive the BPS idea taking
stock of these recent advances.

As in the BPS model we shall use the cusp of a
superconducting string
as the central engine in GRB. It provides the tremendous engine
energy naturally beamed.  Our main observation is that putting
superconducting cusps in a different enviroment, the magnetized
plasma at a relatively small redshift $z$, results in a different
mechanism of
gamma radiation, which leads to a good agreement with GRB
observational data.

GRB radiation in our model arises  as follows.
Low-frequency electromagnetic radiation
from a cusp
loses its energy by accelerating particles of the plasma
to very large Lorentz factors.
Like the initial electromagnetic pulse, the particles are beamed
and give rise to
a hydrodynamical flow in the surrounding gas, terminated by a shock,
as in the standard fireball theory of GRB \cite{MeRee92}
(for a review see
\cite{Pir98}).
We shall assume that cosmic magnetic fields were
generated at moderate redshifts ({\em e.g.}, in young galaxies
during the bright phase of their
evolution \cite{ParPee}).
The string symmetry breaking
scale $\eta$ will be the only string parameter used in our
calculations. With reasonable assumptions about the magnitude of
cosmic magnetic fields and the fraction of volume in the Universe that
they occupy,
this parameter is sufficient to account for all main
GRB observational quantities:
the duration $\tau_{GRB}$, the rate of events $\dot N_{GRB}$, and the
fluence $S$.

We begin with the description of some properties of strings, which
will be used below (for a review see \cite{Book}).

 A horizon-size volume at any time $t$
contains a few long strings stretching across the volume and a large
number of small closed loops. The typical length of a loop at
cosmological time $t$ and the loop number density are given by
\beq
l \sim \alpha t, ~~~~  n_l(t)\sim \alpha^{-1}t^{-3}
\label{n}
\eeq
The exact value of the parameter $\alpha$ in (\ref{n}) is not known.
We
shall assume, following \cite{BB}, that $\alpha$ is determined by the
gravitational backreaction, so that
$\alpha\sim k_g G\mu$,
where  $k_g\sim 50$ is a numerical coefficient, $G$ is
Newton's constant, $\mu\sim \eta^2$ is the mass per unit length of
string, and $\eta$ is the symmetry breaking scale of strings.

The loops oscillate and lose their energy,
mostly by gravitational radiation.  For a loop of invariant length $l$
\cite{invlength}, the
oscillation period is $T_l=l/2$ and the lifetime is
$\tau_l\sim l/k_g G\mu$.

An electric field $E$ applied along a superconducting string generates an
electric current.
A superconducting loop of string oscillating in a magnetic field $B$
acts as an {\it ac} generator and develops an {\it ac} current of
amplitude
\beq
J_0\sim e^2Bl.
\label{J}
\eeq
The local value of the current in the loop can be greatly enhanced in
near-cusp regions where, for a short period of time, the string
reaches a speed very
close to the speed of light.  Cusps tend to be formed a few times
during each oscillation period.  Near a cusp, the string gets
contracted by a large factor, its rest energy being turned into
kinetic energy.  The density of charge carriers, and thus the current,
are enhanced by the same factor.  The contraction factor increases as
one approaches the point of the cusp.

The growth of electric current near the cusp due to string contraction is
terminated at a critical value $J_{max}$ when the energy of charge
carriers becomes comparable to that of the string itself, $(J/e)^2\sim
\mu$.  This gives
\beq
J_{max}\sim e\eta, ~~~~\gamma_{max}\sim (e\eta/J_0).
\label{Jmax}
\eeq
Alternatively, the cusp development can be terminated by small-scale
wiggles on the string \cite{Carter}.  If the wiggles contribute a
fraction $\epsilon\ll 1$ to the total energy of the string, then the
maximum Lorentz factor is less than (\ref{Jmax}), and is given by
$\gamma_{max}\sim\epsilon^{-1/2}$.
The actual value of $\gamma_{max}$ is not
important for most of the following discussion.

Due to the large current, the cusp produces a
powerful pulse of electromagnetic radiation. The total energy of the pulse
is given by \cite{VV,SPG} $ {\cal E}_{em}^{tot} \sim 2 k_{em}J_0 J_{max}l$,
where $l \sim \alpha t$ is the length of the loop, and the coefficient
$k_{em} \sim 10$
is taken from numerical calculations \cite{VV}.
This radiation is emitted within a very narrow cone of
openening  angle $\theta_{min} \sim 1/\gamma_{max}$ (for relativistic
beaming in GRB see \cite{Mao}).
The angular distribution of radiated
energy at larger angles is given by \cite{VV}
\beq
d{\cal E}_{em}/d\Omega \sim k_{em}J_0^2 l /\theta^3,
\label{Emax}
\eeq

We shall adopt the following simple model of cosmic magnetic
fields.  We shall assume that magnetic fields were generated at some
$z\sim z_B$ and then remained frozen in the extragalactic plasma, with
\beq
B(z)=B_0(1+z)^2,
\label{B}
\eeq
where $B_0$ is the characteristic field strength at the present time.
For numerical estimates below we shall use $z_B\sim 4$, $B_0\sim
10^{-7}$ G, and assume that the fraction of volume of the
universe occupied
by magnetized plasma is $f_B\sim 0.1$.  We shall also assume
that the universe is spatially flat and is dominated by
non-relativistic matter.

We shall now estimate the physical quantities characterizing GRBs
powered by superconducting strings.
For a GRB originating at redshift $z$ and seen at angle $\theta$ with
respect to the string velocity at the cusp, we have from
Eqs.(\ref{J})-(\ref{B})
\beq
d{\cal E}_{em}/d\Omega
\sim k_{em}e^4\alpha^3 t_0^3 B_0^2 (1+z)^{-1/2}\theta^{-3},
\label{Eobs}
\eeq
where
$t_0$ is the present age of
the Universe.  The Lorentz factor of the
relevant string segment near the cusp is $\gamma \sim 1/\theta$.
The duration of the cusp event as seen by a distant observer is \cite{BPS}
\beq
\tau_c\sim (1+z)(\alpha t/2)\gamma^{-3}
\sim (\alpha t_0/2) (1+z)^{-1/2}\theta^{3}.
\label{tau}
\eeq
One can expect that the observed duration of GRB is
$\tau_{GRB}\sim\tau_c$.  This expectation will be justified by the
hydrodynamical analysis below.

The fluence, defined as the total energy
per unit area of the detector, is \cite{Paczynski}
\beq
S\sim (1+z)(d{\cal E}_{em}/d\Omega) d_L^2(z),
\label{S}
\eeq
where $d_L(z)=3t_0(1+z)^{1/2}[(1+z)^{1/2}-1]$ is the luminosity
distance.

The rate of GRBs originating at cusps in the redshift interval $dz$
and seen at an angle $\theta$ in the interval $d\theta$ is given by
\beq
d{\dot N_{GRB}}\sim f_B\cdot {1\over{2}}\theta d\theta
(1+z)^{-1}\nu(z) dV(z).
\label{dN}
\eeq
Here, $\nu(t)\sim n_l(t)/T_l\sim 2\alpha^{-2}t^{-4}$ is the number of
cusp events per unit spacetime volume, $T_l\sim \alpha
t/2$
is the oscillation period of a loop,
$dV=54\pi t_0^3[(1+z)^{1/2}-1]^2(1+z)^{-11/2} dz$ is the proper volume between
redshifts $z$ and $z+dz$,
and we have used the relation $dt_0=(1+z)dt$.

Since different cusp events originate at different redshifts and are
seen at different angles, our model automatically gives a distribution
of durations and fluences of GRBs.  The angle $\theta$ is related to
the Lorentz factor of the relevant portion of the string as
$\theta\sim 1/\gamma$, and
from Eqs.(\ref{Eobs}),(\ref{S}) we have
\beq
\gamma (z;S) \sim
\gamma_0 \alpha^{-1}_{-8}S_{-8}^{1/3}B_{-7}^{-2/3}
[(\sqrt{1+z}-1)^2\sqrt{1+z}]^{1/3}.
\label{gamma}
\eeq
Here, $\gamma_0 \approx 190,~~\alpha_{-8}=\alpha/10^{-8}$, and the
fluence $S$ and the magnetic field $B_0$ are expressed
as $S=S_{-8}\cdot 10^{-8}~erg/cm^2$ and $B=B_{-7}\cdot 10^{-7}~G$.

Very large values of $\gamma\sim\gamma_{max}$, which correspond
(for a given redshift) to largest fluences, may not be seen at all because
the radiation is emitted into a too narrow solid angle and the observed
rates of these events are too small.
The minimum value $\gamma(z;S_{min})$
is determined by the
smallest fluence that is observed, $S_{min}\sim 2\cdot
10^{-8}~erg/cm^2$.
Another limit on $\gamma$, which dominates at small $z$, follows
from the condition of compactness \cite{Pir98} and is given by $\gamma
\gtrsim 100$ (see below).


The total rate of GRBs with fluence larger than $S$ is obtained by
integrating Eq.(\ref{dN})
over $\theta$ from
$\gamma_{max}^{-1}(z)$ to $\gamma^{-1}(z;S)$ and over $z$ from $0$
to $\min[z_m;z_B]$, with $z_m$ from $\gamma_{max}(z_m)=\gamma(z_m;S)$.
For relatively small fluences,
$S_{-8}<S_c=0.03(\gamma_{max}(0)\alpha_{-8}/\gamma_0)^3B_{-7}^2$,
$z_B<z_m$
and we obtain
\begin{eqnarray}
{\dot N_{GRB}(>S)} &\sim& \frac{f_B}{2\alpha^2 t_0^4 }
\int_0^{z_B}dV(z)(1+z)^5\gamma^{-2}(z;S) \nonumber\\
&\sim& 3\cdot 10^2 S_{-8}^{-2/3}B_{-7}^{4/3}~yr^{-1}.
\label{rate}
\end{eqnarray}
Remarkably, this rate in our model does not depend on any string
parameters and is
determined (for a given value of $S$) almost entirely by the
magnetic field $B_0$.
The predicted slope $\dot N_{GRB}(>S) \propto S^{-2/3}$
is in a reasonable agreement with the observed one
${\dot N}_{obs}(>S)\propto S^{-0.55}$ at relatively small fluences
\cite{Bie99}.

For large fluences $S_{-8} >S_c$,
integration of Eq.(\ref{dN}) gives
$\dot N_{GRB}(>S) \propto S^{-3/2}$. Observationally, the transition
to this regime occurs at $S_{-8}\sim 10^2-10^3$.  This can be
accounted for if the cusp
development is terminated by small-scale wiggles with fractional
energy in the wiggles
{\bf $\epsilon\sim 10^{-7}\alpha_{-8}^2 B_{-7}^{4/3}$}.
Alternatively, if $\gamma_{max}$ is
determined by the back-reaction of the charge carriers,
Eq.(\ref{Jmax}), then the regime
(\ref{rate}) holds for larger $S_{-8}$, and observed steepening of the
distribution at large $S$ can be due to the reduced efficiency of BATSE
to detection of
bursts with large $\gamma$. Indeed, large $\gamma$ results in
a large Lorentz factor $\gamma_{CD}$ of the emitting region (see below), and
at $\gamma_{CD} \gtrsim 10^3$ photons start to escape from the
BATSE range.

The duration of a GRBs originating at redshift $z$ and having fluence
$S$ is readily calculated as
\beq
\tau_{GRB} \approx 200 \frac{\alpha_{-8}^4 B_{-7}^2}{S_{-8}}
(1+z)^{-1}(\sqrt{1+z}-1)^{-2}~s
\label{duration}
\eeq

Estimated from Eq.(\ref{tau}),  $\tau_{GRB}^{max} \sim 10^3\alpha_{-8}~s$,
while from Eq.(\ref{duration}) using
$S_{max} \sim 1\cdot 10^{-4}~erg/cm^2$ and $z \sim z_B\sim 4$,
one obtains $\tau_{GRB}^{min}\sim 3\alpha_{-8}^4 B_{-7}^2~ms$. This
range of $\tau_{GRB}$ agrees with observations for
$\alpha_{-8} \sim 1$  ({\em i.e.}
$\eta \sim 2\cdot 10^{14}~GeV$) and $B_{-7} \sim 1$.

Let us now turn to the hydrodynamical phenomena in which the gamma radiation
of the burst is actually generated.
The low-frequency electromagnetic pulse
interacting with
surrounding gas produces an ultrarelativistic
beam of accelerated particles. This is the dominant channel of energy
loss by the pulse. The beam of high energy particles pushes the
gas with the frozen magnetic field ahead of it, producing an external
shock in surrounding plasma  and  a reverse  shock
in the beam material, as in the case of ``ordinary'' fireball
(for a review see \cite{Pir98}). The difference is that the beam propagates
with a very large Lorentz factor $\gamma_b \gg \gamma$
where $\gamma$ is the Lorentz factor of the cusp.
(The precise value of
$\gamma_b$ is not important for this discussion; it will be estimated
in a subsequent publication \cite{BHV2}.)  Another difference is
that the beam
propagates in a very low-density gas. The beam can be regarded as a
narrow shell of relativistic particles of width $\Delta \sim l/2\gamma^3$
in the observer's frame.

The gamma radiation of the burst
is produced as synchrotron radiation of electrons accelerated by
external and reverse shocks. Naively, the duration of synchrotron
radiation, i.e. $\tau_{GRB}$, is determined by the thickness of the
shell as $\tau_{GRB} \sim \Delta$.  This is confirmed by a more
detailed analysis, as follows.
The reverse shock in our case is ultrarelativistic
\cite{KobPirSar98,Pir98}.
The  neccessary
condition for that, $\rho_b/\rho < \gamma_b^2$, is satisfied with a
wide margin
(here $\rho_b$ is the baryon density in the beam and $\rho$ is the
density of unperturbed gas).  In this case,
the shock dynamics and the GRB duration are determined by two
hydrodynamical parameters \cite{Pir98}.
They are the thickness of the shell $\Delta$
and the Sedov length, defined as the distance travelled by the shell
when the mass of the snow-ploughed
gas becomes comparable to the initial energy of the beam. The latter is
given by $l_{Sed}\sim({\cal E}_{iso}/\rho)^{1/3}$.

The reverse shock enters the shell and, as it propagates there, it strongly
decelerates the shell. The synchrotron radiation occurs mainly in
the shocked regions of the shell and of the external plasma.
The surface separating these two regions,
the contact discontinuity (CD) surface,  propagates with
the same velocity as the shocked plasma, where the GRB radiation is
produced.

The synchrotron radiation ceases when the reverse shock reaches the
inner boundary of the shell. This occurs at a distance
$R_{\Delta} \sim l_{Sed}^{3/4}\Delta^{1/4}$
when the Lorentz
factor of the CD surface is $\gamma_{CD}\sim (l_{Sed}/\Delta)^{3/8}
\sim 0.1 B_{-7}^{1/4}n_{-5}^{-1/8}(1+z)^{1/2}\gamma^{3/2}$, where
$n_{-5}$ is the ambient baryon number density in units of $10^{-5}$
cm$^{-3}$.  Note that these
values do not depend on  the Lorentz factor of the beam $\gamma_b$
and are determined by the cusp Lorentz factor $\gamma$.
The size of the synchrotron emitting region is of the order $R_{\Delta}$,
and the Lorentz factor of this region is equal to $\gamma_{CD}$.
The compactness condition \cite{Pir98} requires $\gamma_{CD}\gtrsim
10^2$, which yields $\gamma\gtrsim 10^2$ used above.
The duration of GRB is given by
\beq
\tau_{GRB}\sim {R_\Delta}/{2\gamma_{CD}^2}\sim {l}/{2\gamma^3},
\label{taugrb}
\eeq
{\em i.e.} it is equal to the duration of the cusp event given by
Eq.(\ref{tau}).
The energy that goes into
synchrotron radiation is comparable to the energy of the
electromagnetic pulse.

We conclude with several remarks on our GRB model.

{\it (i)} Magnetic fields in our scenario could be generated in young
galaxies during the bright phase of their evolution \cite{ParPee} and then
dispersed by galactic winds in the intergalactic space.  One expects
that at present the fields are concentrated in the filaments and
sheets of the large-scale structure \cite{Bier,param}.  With sheets of
characteristic size $L\sim (20-50)h^{-1}$ Mpc and thickness $D\sim
5h^{-1}$ Mpc, we have $f_B\sim D/L\sim 0.1$.  The average field
strength can be estimated  from the equipartition condition as
$B_0\sim 10^{-7}$ G \cite{Bier}.  We expect these values to be accurate
within an order of magnitude.

{\it (ii)} We emphasize that our model involves a number of
simplifying assumptions.  All loops at cosmic time $t$ were assumed to
have the same length $l\sim\alpha t$, while in reality there should be
a distribution $n(l,t)$.  The evolution law (\ref{B}) for $B(z)$ and
the assumption of $f_B={\rm const}$ are also oversimplified.  A more
realistic model should also account for a spatial variation of $B$.
As a result, the correlation between $\tau$ and $S$ suggested by
Eq.(\ref{duration}) will tend to be washed out.  Finally, we disregarded
the possibility of a nonzero cosmological constant which would
introduce some quantitative changes in our estimates.

{\it (iii)} Apart from the reverse shock instability,
small-scale wiggles on strings can naturally produce short-time
variation in GRBs. These wiggles, acting like mini-cusps, produce a
sequence of successive fireballs, which are the usual explanation of
GRB short-time structure \cite{Pir98}.

{\it (iv)} Cusps reappear on a loop, producing nearly identical
GRBs with a period $T_l \sim \alpha t/2 \sim 50(1+z)^{-1/2}~yr$.
In a more realistic model, some fraction of loops would have
lengths smaller than $\alpha t$ and shorter recurrence periods.

{\it (v)} Our model predicts that GRBs should be accompanied by
strong bursts of gravitational radiation.  The angular distribution of
the gravitational wave energy around the direction of the cusp is
\cite{VV84} $d{\cal E}_g/d\Omega\sim G\mu^2/\theta$, and the
dimensionless amplitude of a burst of duration $\tau$
originating at redshift $z$ can be estimated as
\beq
h\sim k_g(G\mu)^2(\tau/\alpha t_0)^{1/3}(1+z)^{-1/3}[(1+z)^{1/2}-1]^{-1},
\eeq
or $h\sim 10^{-21}\alpha_{-8}^{5/3}z^{-1}(\tau/1s)^{1/3}$ for $z\lesssim
1$.  Here, we have used the relation $F_g\sim h^2/G\tau^2 \sim (1+z)
(d{\cal E}_g/d\Omega) d_L^{-2}(z)$ for the gravitational wave flux and
Eq.(\ref{tau}) for the burst duration $\tau$.
These gravitational wave bursts are much stronger
than expected from more conventional
sources and should be detectable by the planned LIGO, VIRGO and LISA
detectors.

{\it (vi)} GRBs have been suggested as possible sources of the
observed ultrahigh-energy cosmic rays (UHECR) \cite{Vie,Wax}.
This idea encounters two difficulties. If GRBs are
distributed uniformly in the universe, UHECR have
a classical  Greisen-Zatsepin-Kuzmin (GZK)
cutoff, absent in the observations. The acceleration by an
ultrarelativistic shock is possible only in the one-loop regime ({\em i.e.}
due to a single reflection from the shock) \cite{GalAch}.
For a standard GRB with a
Lorentz factor $\gamma_{sh} \sim 300$ it results in the maximum energy
$E_{max} \sim \gamma_{sh}^2 m_p \sim 10^{14}~eV$,  far too low for UHECR.

Our model can resolve both of these difficulties, assuming that
$gamma_{max}$ is determined by the current backreaction, Eq.(\ref{Jmax}).

If the magnetic field in the Local Supercluster (LS) is considerably
stronger than outside, then the cusps in LS
are more powerful and the GZK cutoff is less pronounced.

Near-cusp segments with large Lorentz factors produce
hydrodynamical flows with large Lorentz factors, {\em e.g.}
$\gamma \sim 2\cdot 10^4$ corresponds to $\gamma_{CD} \sim 3\cdot
10^5$ and
$E_{max} \sim \gamma_{CD}^2m_p \sim 1\cdot 10^{20}~eV$. Protons with
such energies are deflected in the magnetic field of LS and can be
observed, while protons with much higher energies caused by near-cusp
segments with $\gamma\gtrsim 10^5$ are propagating rectilinearly and
generally are not seen.  Further details of this scenario
will be given elsewhere \cite{BHV2}.

We are grateful to Ken Olum for useful discussions.  The work of AV
was supported in part by the National Science Foundation and the work
of VB and BH by INTAS through grant No 1065.


\begin{thebibliography}{99}
\bibitem{invlength} The actual length of the loop changes as the loop
oscillates.  The invariant length is defined as $l=E/\mu$, where $E$
is the loop's energy in its center-of-mass frame.
\end{thebibliography}
\end{document}